\documentclass[twocolumn,preprintnumbers,amsmath,amssymb,floatfix]{revtex4}

\usepackage{graphicx}
\usepackage{dcolumn}
\usepackage{bm}

\begin{document}

\title{Centrality dependence of charged hadron transverse momentum spectra in d+Au collisions 
at $\sqrt{s_{_{\it NN}}} =$ 200 GeV }
\author{
%
%
B.B.Back$^1$,
M.D.Baker$^2$,
M.Ballintijn$^4$,
D.S.Barton$^2$,
B.Becker$^2$,
R.R.Betts$^6$,
A.A.Bickley$^7$,
R.Bindel$^7$,
A.Budzanowski$^3$,
W.Busza$^4$,
A.Carroll$^2$,
M.P.Decowski$^4$,
E.Garc\'{\i}a$^6$,
T.Gburek$^3$,
N.George$^{1,2}$,
K.Gulbrandsen$^4$,
S.Gushue$^2$,
C.Halliwell$^6$,
J.Hamblen$^8$,
A.S.Harrington$^8$,
C.Henderson$^4$,
D.J.Hofman$^6$,
R.S.Hollis$^6$,
R.Ho\l y\'{n}ski$^3$,
B.Holzman$^2$,
A.Iordanova$^6$,
E.Johnson$^8$,
J.L.Kane$^4$,
N.Khan$^8$,
P.Kulinich$^4$,
C.M.Kuo$^5$,
J.W.Lee$^4$,
W.T.Lin$^5$,
S.Manly$^8$,
A.C.Mignerey$^7$,
A.Noell$^7$,
R.Nouicer$^{2,6}$,
A.Olszewski$^3$,
R.Pak$^2$,
I.C.Park$^8$,
H.Pernegger$^4$,
C.Reed$^4$,
L.P.Remsberg$^2$,
C.Roland$^4$,
G.Roland$^4$,
J.Sagerer$^6$,
P.Sarin$^4$,
P.Sawicki$^3$,
I.Sedykh$^2$,
W.Skulski$^8$,
C.E.Smith$^6$,
P.Steinberg$^2$,
G.S.F.Stephans$^4$,
A.Sukhanov$^2$,
R.Teng$^8$,
M.B.Tonjes$^7$,
A.Trzupek$^3$,
C.Vale$^4$,
G.J.van~Nieuwenhuizen$^4$,
R.Verdier$^4$,
G.I.Veres$^4$,
B.Wadsworth$^4$,
F.L.H.Wolfs$^8$,
B.Wosiek$^3$,
K.Wo\'{z}niak$^3$,
A.H.Wuosmaa$^1$,
B.Wys\l ouch$^4$,
J.Zhang$^4$\\
\vspace{2mm}
(PHOBOS Collaboration) \\
\vspace{2mm}
\small
%
%
%
%
$^1$~Argonne National Laboratory, Argonne, IL 60439-4843, USA\\
$^2$~Brookhaven National Laboratory, Upton, NY 11973-5000, USA\\
$^3$~Institute of Nuclear Physics, Krak\'{o}w, Poland\\
$^4$~Massachusetts Institute of Technology, Cambridge, MA 02139-4307, USA\\
$^5$~National Central University, Chung-Li, Taiwan\\
$^6$~University of Illinois at Chicago, Chicago, IL 60607-7059, USA\\
$^7$~University of Maryland, College Park, MD 20742, USA\\
$^8$~University of Rochester, Rochester, NY 14627, USA\\
}

\date{\today}

\begin{abstract}\noindent

We have measured transverse momentum distributions of charged hadrons 
produced in d+Au collisions at $\sqrt{s_{_{\it NN}}} =$ 200 GeV. 
The spectra were obtained for transverse momenta $0.25 < p_T < 6.0$~GeV/c, in a 
pseudorapidity range of $0.2 < \eta < 1.4$ in the deuteron direction. 
The evolution of the spectra with collision centrality is presented in comparison
to $p+\bar{p}$ collisions at the same collision energy.  
With increasing centrality, the yield at high transverse momenta 
increases more rapidly than the overall particle density, leading to a strong modification
of the spectral shape. This change in spectral shape is qualitatively different from 
observations in Au+Au collisions at the same energy. The results provide important information
for discriminating between different models for the suppression of high-$p_T$ hadrons observed
in Au+Au collisions.

\vspace{3mm}
\end{abstract}

\maketitle

The yield of charged hadrons produced in collisions of deuterons with gold nuclei at 
an energy of $\sqrt{s_{_{\it NN}}} = 200$~GeV has been measured 
with the PHOBOS detector at the Relativistic Heavy Ion Collider (RHIC)
at Brookhaven National Laboratory.
The data are  presented as a function of transverse momentum ($p_T$) and collision centrality.
The goal of these measurements is to study the modification
of particle production due to initial state effects in the nuclear medium,
in comparison to nucleon-nucleon collisions at the same energy.
Measurements from proton-nucleus (p+A) reactions at lower collision energies
have found that the cross-section for hadron production at $p_T$ of 1.5 to
5 GeV/c rises faster than the nuclear size $A$ \cite{cronin}.  
This observation, which is commonly called the Cronin effect,
has been described as the result of initial state multiple scattering, leading to a
broadening of the $p_T$ distribution \cite{accardi}.

The present interest in repeating these measurements at higher energies 
is motivated by  results from Au+Au collisions at $\sqrt{s_{_{\it NN}}} =$~130 and 200~GeV.
In these collisions, the expected scaling of hadron production with the 
number of binary nucleon-nucleon collisions at $p_T$ of 2--10 GeV/c 
is strongly violated 
\cite{phenix_quench,phenix_highpt_npart,star_highpt_npart, phobos_highpt_npart}. 
This effect had been predicted as a consequence of the energy loss of high-$p_T$ partons
in the hot and dense medium formed in Au+Au collisions \cite{jet_quench_theory}.
The interpretation of the Au+Au data relies on the understanding 
of initial state effects, including gluon saturation \cite{kharzeev}, 
which can be investigated with the d+Au data presented here \cite{wang_dAu}. Similar measurements are
reported in \cite{phenix_dAu,star_dAu}.
By studying the spectra as a function of collision centrality, we can 
control the effective thickness of nuclear matter traversed by the incoming partons.

The data were collected using the PHOBOS 
two-arm magnetic spectrometer \cite{phobos_nim}. 
The spectrometer arms 
are each equipped with  16 layers of silicon sensors, providing 
charged particle tracking both outside and 
inside the 2~T field of the PHOBOS magnet. 
Additional silicon detectors used in this analysis are the central single-layer
Octagon barrel detector and the three single-layer forward Ring detectors 
located on either side of the interaction point.

The primary event trigger (Level 0) was provided by two sets of 16 scintillator 
counters (``Paddle counters'') covering pseudorapidities 
$3 < |\eta |< 4.5$. In addition, two higher level trigger conditions were used.
Collisions close to the nominal vertex position $z_{vtx} = 0$ along the longitudinal ($z$) direction were 
selected using the time difference between signals in 
two rings of ten \v Cerenkov counters. The counters covered
$-4.9 < \eta < -4.4$ and $3.6 < \eta < 4.1$, respectively.
For part of the data set, a further online selection of events 
was accomplished using two arrays of horizontally segmented scintillator hodoscopes. 
One array was positioned immediately behind the spectrometer and the other at a 
distance of 
$\sim 5$~m from the interaction point.  In combination with the known vertex 
position, a spatial coincidence from the hodoscopes was used to 
trigger on high-$p_T$ particles which traverse the spectrometer.

Previous measurements of p+A collisions have shown a strong
dependence of the multiplicities and momentum distributions of produced particles 
on the size of the target nucleus \cite{cronin,halliwell}. For d+Au collisions, 
this implies the importance of characterizing 
the collision centrality. For a given event selection, the centrality can be quantified 
by a number of variables such as the average number of participating nucleons 
$\langle N_{\it part} \rangle$ or the average number 
of binary nucleon-nucleon collisions, $\langle N_{\it coll}\rangle$. Neither of these 
variables can be directly measured. 
Estimates of $\langle N_{\it part} \rangle$ or 
$\langle N_{\it coll}\rangle$ for a given event selection are typically obtained by matching the distribution of 
an experimental observable, such as the multiplicity in a certain $\eta$-region, 
with the results of a Glauber model calculation and detector simulation \cite{phobos_cent_200}. 

\begin{table}[htbp]
\begin{center}
\begin{tabular}{|c|c|c|c|}
\hline Centrality Selection & $ \left<N_{\it part}\right>$ & $\left<N_{\it coll}\right>$ & Efficiency \\
\hline
 0--20\%   & $15.5  \pm 1.0$   & $14.6 \pm 0.9$ &  82\% \\
 20--40\%  & $10.9  \pm 0.9$   & $9.7  \pm 0.8$ &  73\% \\
 40--70\%  & $6.7   \pm 0.9$   & $5.4  \pm 0.8$ &  49\% \\
 70--100\% & $3.3   \pm 0.7$   & $2.2  \pm 0.6$ &  14\% \\
\hline
\end{tabular}
\caption{
\label{table1}
Estimated values for $\left<N_{\it part}\right>$ and  $\left<N_{\it coll}\right>$ and the respective systematic 
uncertainties for the four centrality bins used in this analysis. Also shown is the average event 
selection efficiency for each of the bins. The centrality bins are based on 
the signal of multiplicity counters covering $3.0 < |\eta| < 5.4$.}
\end{center}  
\end{table}

The centrality cuts for this analysis were based on the signal ($E_{\it Ring}$) of the three Ring 
detectors in the region of $3.0 < |\eta | < 5.4$, which is proportional to the 
number of charged particles hitting these counters. In the MC simulations, centrality cuts on $E_{\it Ring}$ were found to 
introduce a bias on the yield in the spectrometer acceptance of less than 5\% for $\langle N_{\it part} \rangle> 3$,
compared to cutting directly on  $N_{\it part}$. 
If the $N_{\it part}$ determination is based on a multiplicity region closer to
the spectrometer acceptance, MC studies, as well as our data, showed 
a significantly larger bias at low and high centralities. 
For each of the four bins in $E_{\it Ring}$, $\langle N_{part} \rangle$ and $\langle N_{coll} \rangle$ were obtained from a 
Glauber model calculation using HIJING 
\cite{hijing} and a full detector simulation which  included the experimental trigger and event selection
efficiency. 
In the HIJING calculations, the default value for the inelastic 
nucleon-nucleon cross-section of 41~mb was used, 
consistent with previous calculations \cite{phobos_cent_200}.

The resulting estimates for $\langle N_{\it part} \rangle$ and $\langle N_{\it coll}\rangle$ 
in the four $E_{\it Ring}$-based centrality bins are shown in Table~I.
The percentage numbers refer to the fractional cross-section in the unbiased HIJING distribution. 
The determination of $\langle N_{\it part} \rangle$ and $\langle N_{\it coll}\rangle$ takes into 
account the bias introduced in the centrality by the online and offline event selection 
in the relatively low-multiplicity d+Au events.
The largest contribution to this bias comes from the online vertex trigger, leading to the 
average event selection efficiencies for the individual centrality bins shown 
in Table~I. 
\begin{figure}[htbp]
\includegraphics[width=7cm]{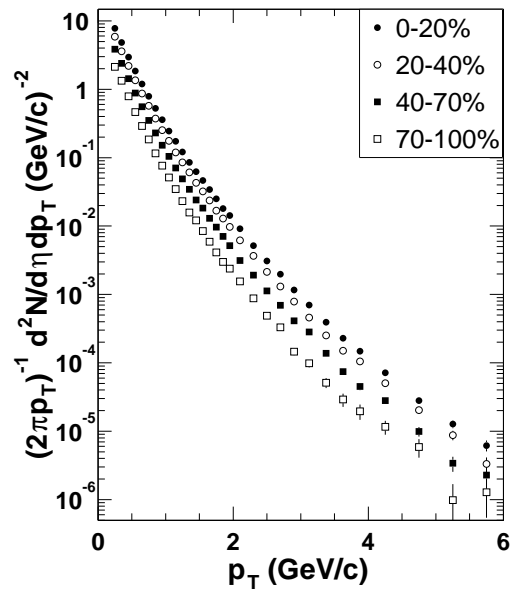}
\caption{ \label{ref_SpectraAllCent} 
Invariant yield of charged hadrons, $\frac{h^+ + h^-}{2}$, as a function of $p_T$ for four
centrality bins.  Only statistical errors are shown.}
\end{figure}

Details of the track reconstruction algorithm can be found in \cite{pbarp_200,phobos_highpt_npart}.
To optimize the momentum resolution and minimize systematic errors in the track selection,
only particles traversing the full spectrometer arms were included in the analysis.
These particles leave a minimum of 12 hits in the silicon detectors. This selection limits
the usable vertex range to $-15~\mbox{cm}~< z_{vtx}~<~+10$~cm.
Due to the low multiplicity in d+Au collisions, a new algorithm for the offline determination of the 
collision vertex was developed, using hit position and energy information in the Octagon detector. MC studies
show a resolution in the beam direction of $\sigma_{vtx_{z}} = 1.4$~cm for the most peripheral 
events and $\sigma_{vtx_{z}} = 0.8$~cm for the most central events. The transverse position of the 
event vertex was centered at the known position of the beam orbit. Unlike in the Au+Au track finding,
the vertex position information was not included in the initial track seed.
\begin{figure}[t]
\includegraphics[width=7cm]{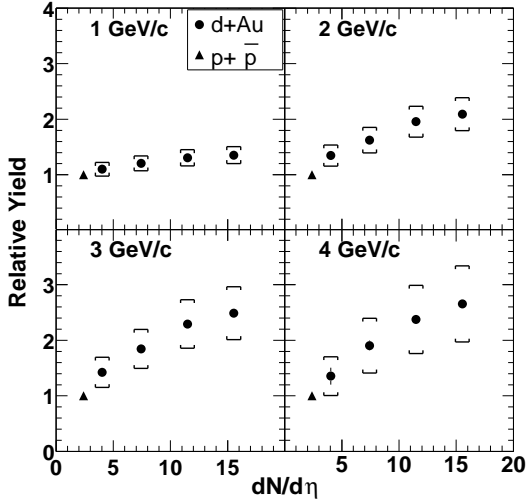}
\caption{ \label{ref_CroninVsYield}
The relative yield  of d+Au to UA1 $p+\bar{p}$ data at $p_T = 1, 2, 3 \mbox{ and } 4$~GeV/c is shown
as a function of the integrated yield $dN/d\eta$ for the four $E_{\it Ring}$ centrality bins. 
The ratio d+Au/$p+\bar{p}$ has been normalized to unity at $p_T = 0.5$ GeV/c. The triangles 
indicate the values for $p+\bar{p}$. 
The brackets indicate the systematic errors on the relative yield (90\% C.L.). The systematic
error on $dN/d\eta$ is 12\%.} 
\end{figure}

To obtain the invariant yield of charged hadrons, we accumulated equal amounts of data with both
magnet polarities. The transverse momentum
distributions for each centrality bin were corrected for the geometrical acceptance of the detector, the
efficiency of the tracking algorithm and the distortion due to binning and momentum
resolution. The procedure for obtaining the correction factors was described in 
\cite{phobos_highpt_npart}.
The largest contributions to the systematic uncertainty come from the overall tracking 
efficiency (5--10\% uncertainty)
and the reduction in overall acceptance due to malfunctioning channels in the 
silicon detectors (5\% uncertainty). 
The corresponding corrections are centrality independent. The next largest correction is 
the $p_T$ and centrality dependent momentum resolution and binning 
correction. The contamination by secondary particles and feeddown particles is small, due to the 
proximity of the tracking detectors to the collision vertex and the requirement for the 
reconstructed track to point back to the beam orbit to  within 0.4~cm.

\begin{figure}[t]
\includegraphics[width=8.5cm]{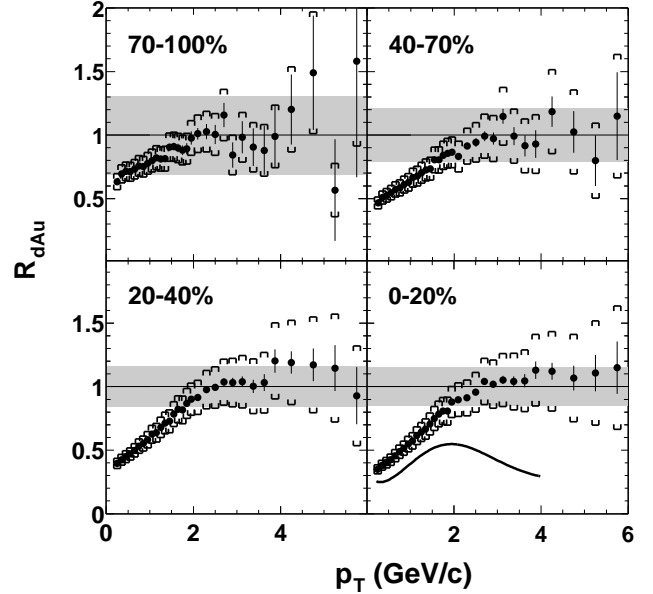}
\caption{ \label{ref_UA1Compare} 
Nuclear modification factor $R_{\it dAu}$ as a function of $p_T$ for four 
bins of centrality. 
For the most central bin, the spectral shape for central Au+Au data relative to $p+\bar{p}$ is shown for comparison.
The shaded area shows the uncertainty in $R_{\it dAu}$ due to the systematic uncertainty in 
$\langle N_{\it coll} \rangle$ and the UA1 scale error (90\% C.L.). 
The brackets show the systematic uncertainty of the d+Au spectra measurement (90\% C.L.).}
\end{figure}

In Fig.~\ref{ref_SpectraAllCent}, we present the invariant yield
of charged hadrons as a function of transverse momentum, obtained by
averaging the yields of positive and negative hadrons. 
Data are shown for four $E_{\it Ring}$ centrality bins. The plot shows 
the evolution of overall yield and spectral shape with increasing collision centrality.

The centrality evolution of the spectra can be studied in detail in Fig.~\ref{ref_CroninVsYield}, where
we compare our d+Au data to results from  UA1 for $p+\bar{p}$ collisions at the same energy \cite{ua1_pbarp}.
To account for the difference in acceptance between UA1 ($|\eta| < 2.5$) and PHOBOS, a correction 
function was determined using PYTHIA \cite{pythia}. 
The quantity shown on the vertical axis in Fig.~\ref{ref_CroninVsYield} is a direct measure of the modification
of the spectral shape relative to $p+\bar{p}$ for each
centrality bin in d+Au and is defined as follows: the fitted $dN/dp_T$ distribution for each centrality bin
is divided by the corrected fit to the  UA1 $p+\bar{p}$ data \cite{fit}. The resulting distribution is normalized to 
unity at $p_T = 0.5$~GeV/c. Then the value of the normalized ratio at four values
of $p_T$ from 1 to 4 GeV/c is plotted against $dN/d\eta$ determined
by integrating the hadron spectrum for each centrality bin. If there were no modification of the 
spectral shape relative to $p + \bar{p}$ collisions, the ratio would be flat at unity. 
This presentation of the data allows an investigation of the evolution of the spectral shape
with increasing centrality, while eliminating the uncertainty associated with the determination
of $\langle N_{\it coll} \rangle$.
We observe that the relative yield for all $p_T$ regions grows smoothly as a function of $dN/d\eta$,
with the biggest increase relative to $p+\bar{p}$ observed at the largest $p_T$. For all $p_T$ regions,
the data extrapolate to the corrected $p+\bar{p}$ fit.

\begin{figure}[t]
\includegraphics[width=7cm]{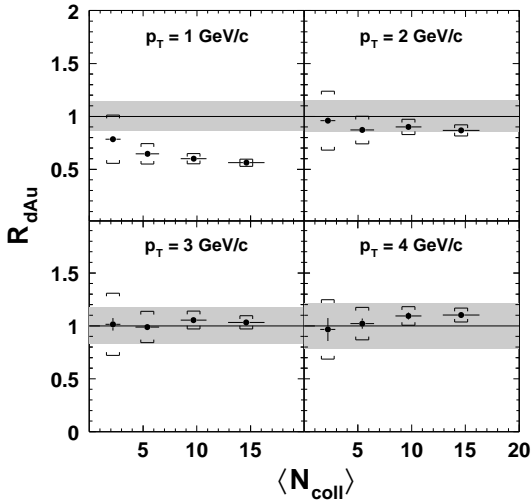}
\caption{ \label{ref_RdAu}
Nuclear modification factor $R_{\it dAu}$ as a function of centrality in four bins of transverse
momentum. The brackets indicate the point-to-point systematic error, dominated by the uncertainty
in the number of collisions for each centrality bin. The grey band shows the overall scale uncertainty
at each $p_T$. Systematic errors are at 90\% C.L.}
\end{figure}

In Fig.~\ref{ref_UA1Compare} we present the nuclear modification factor $R_{\it dAu}$ as a function
of $p_T$ for each centrality bin, defined as 
\begin{equation}
R_{\it dAu} = \frac{\sigma_{p\overline{p}}^{inel}}{\langle N_{\it coll} \rangle} 
              \frac{d^2 N_{\it dAu}/dp_T d\eta} {d^2 \sigma(\mbox{UA1})_{p\overline{p}}/dp_T d\eta}.
\end{equation}
Consistent with our Glauber calculations, we used $\sigma_{pp}^{inel} = 41$~mb. A value 
of $R_{dAu} = 1$ corresponds to scaling of the yield as an incoherent superposition
of nucleon-nucleon collisions.
For all centrality bins, we observe a rapid rise of $R_{dAu}$ from low $p_T$, leveling off
at $p_T$ of $\approx 2$~GeV/c. 
For comparison, we also plot the results from central Au+Au collisions
at the same energy \cite{phobos_highpt_npart} in the lower right panel of Fig.~\ref{ref_UA1Compare}.
The average number of collisions undergone by each participating nucleon in the central
Au+Au collision is close to 6, similar to that of each nucleon from the deuteron in a 
central d+Au collision.
For central Au+Au collisions, the ratio of the spectra to $p+\bar{p}$ rises rapidly 
up to $p_T \approx 2$~GeV/c, but falls far short of collision scaling at larger 
$p_T$, in striking contrast to the behavior for central d+Au collisions.

Predictions for the evolution of $R_{\it dAu}$ from semi-peripheral collisions with
$\langle N_{\it coll} \rangle \approx 6$ to central  collisions were made in 
two qualitatively different models. 
Perturbative QCD calculations \cite{vitev} predict an increase in the maximum value 
of $R_{dAu}$ at $p_T \approx 3.5$~GeV/c by 15\%. 
In contrast, a decrease in $R_{\it dAu}$ by 25--30\% over the same centrality range 
is predicted in a parton saturation model \cite{kharzeev}. The centrality evolution
of $R_{dAu}$ is shown in Fig.~\ref{ref_RdAu}, where the points were obtained from
a fit to the $p_T$ dependence of $R_{dAu}$ in each centrality bin. Our 
data disfavor the prediction from the parton saturation model. This suggests
that the observed suppression of high $p_T$ hadrons in Au+Au collisions 
\cite{phenix_quench,phenix_highpt_npart,star_highpt_npart, phobos_highpt_npart} 
cannot be accounted for by initial state effects that should also be present in d+Au collisions.

This work was partially supported by US DoE grants DE-AC02-98CH10886,
DE-FG02-93ER40802, DE-FC02-94ER40818, DE-FG02-94ER40865,
DE-FG02-99ER41099, W-31-109-ENG-38, US NSF grants 9603486, 9722606,
0072204, Polish KBN grant 2-P03B-10323, and NSC of Taiwan contract NSC
89-2112-M-008-024.


\begin{thebibliography}{99}
\bibitem{fit} The reference spectrum is well fitted by 
$\frac{1}{p_T} \frac{dN}{dp_T d\eta} = 50.9 \cdot (1+p_T/1.6)^{-11.143} \cdot \frac{p_T}{\sqrt{p_T^2 + 0.00545}}$, with
$p_T$ given in GeV/c.
\end{thebibliography}
\end{document}